\documentclass[sort&compress,twocolumn,5p]{elsarticle} 

\usepackage{amsmath}

\usepackage[pdftex,breaklinks,
            pdftitle={An assessment of some closed-form expressions for the Voigt function III},
            pdfsubject={Approximations and expansions, Computational techniques},
            pdfkeywords={Plasma dispersion function, Complex probability function, Hjerting function, Pseudo-Voigt function},
            pdfauthor={F. Schreier}]{hyperref}

\newcommand{\qeq}[1]  {Eq.~(\ref{#1})}
\newcommand{\qufig}[1] {Fig.~\ref{#1}}
\newcommand{\D}{\mathrm{d}}

\newcommand{\I}{\mathrm{i}}
\newcommand{\erf} {\mathop{\rm erf}}
\newcommand{\erfc} {\mathop{\rm erfc}}
\newcommand{\erfce} {\mathop{\rm erfce}}
\newcommand{\half}{\tfrac{1}{2}}
\newcommand{\xHalf} {x_\text{h}}

\batchmode

\journal{J.\ Quant.\ Spectros.\ \& Rad.\ Transfer, received December 2018, accepted January 2019; doi: 10.1016/j.jqsrt.2019.01.017}
\date{}

\begin{document}

\begin{frontmatter}

\title{Notes: \\ An assessment of some closed-form expressions for the Voigt function III: \\
       Combinations of the Lorentz and Gauss functions} 

\author{Franz Schreier}
\address{DLR --- German Aerospace Center, 
         Remote Sensing Technology Institute, \\
         Oberpfaffenhofen, 82234 We\ss ling, Germany}
\cortext[ca]{Corresponding author} 
\ead{franz.schreier@dlr.de}

\begin{abstract}
A variety of ``pseudo-Voigt'' functions, i.e.\ a linear combination of the Lorentz and Gauss function (occasionally augmented with a correction term), have been proposed as a closed-form approximation for the convolution of the Lorentz and Gauss function known as the Voigt function.
First, a compact review of several approximations using a consistent notation is presented.
The comparison with accurate reference values indicates relative errors as large as some percent.
\end{abstract}

\begin{keyword}
Complex error function; Complex probability function; Plasma dispersion function; Faddeyeva function; Pseudo-Voigt function \\
\PACS \\
 02.30.Mv 	Approximations and expansions \\
 02.60.-x 	Numerical approximation and analysis \\
 02.70.-c	Computational techniques \\
\end{keyword}


\end{frontmatter}


\section{Introduction}
\label{sec:intro}

Rapid yet accurate computation of the Voigt function \citep{Armstrong67} is a challenge in many fields of the physical sciences.
Because the convolution integral of a Lorentz and Gauss function does not have an analytical solution, approximations have been discussed in numerous papers.
Whereas many modern ``state-of-the art'' algorithms evaluate the closely related complex error function (also known as complex probability function, plasma dispersion function, or Fadde(y)eva function, cf.\ e.g., \citealp{NIST-handbook,DLMF,Oldham09}) utilizing sophisticated numerical techniques,
``simple'' closed-form expressions still appear to be attractive.

Linear combinations of the Lorentz and/or Gauss functions have been suggested by several authors.
\citet{Flores91} proposed a sum of the Lorentzian and its derivatives and \citet{Melcher77} fitted the Voigt functions by ``generalized Lorentz functions''. 
\citet{McLean94} have further developed an approximation originally suggested by \citet{Martin81,Martin81e,Puerta81,Puerta83e} and proposed a
superposition of four Lorentzians \citep[for a recent assessment see][]{Schreier18v}.
Linear combinations of a Lorentz and Gauss function (sometimes called  ``pseudo-Voigt'' function and occasionally including a correction term)
have been suggested or used by \citet{Whiting68,Matveev72,Peyre72,Kielkopf73,Wertheim74,Thompson87,Teodorescu94,Ida00} and \citet{Liu01}.

In this note we present an assessment of closed-form expressions for the Voigt function using a combination of Lorentz and Gauss functions.
After a short review of the basic definitions in the next subsection, we describe several combinations using a consistent notation (in chronological order).
The results of our tests using an accurate Voigt function code as reference are presented in Section \ref{sec:results}.
The codes have been implemented in Python and a Scientific Python (\url{scipy.org}) implementation has been used as a reference.
In the final Section \ref{sec:conclusions} we provide a summary and some conclusions.


\section{Theory}
\label{sec:theory}

\subsection{The Voigt function}
\label{sec:voigt}

The Voigt function (normalized to $\sqrt\pi$) is defined by 
\begin{equation} \label{vgtFct} 
 K(x,y) ~=~ {y \over \pi} ~ \int_{-\infty}^\infty {\mathrm{e}^{-t^2} \over (x-t)^2 + y^2} \, \D t
\end{equation}
where $x$ is a measure of the distance to the center peak, and $y$ is essentially the ratio of the Lorentzian and Gaussian width, $y = \sqrt{\ln 2} \gamma_\text{L}/\gamma_\text{G}$.
At the line center $x=0$ the Voigt function can be expressed as the exponentially scaled complementary error function
\begin{align} \label{vgtFct0}
 K(0,y) ~=~ \exp(y^2) \bigl( 1 - \erf(y) \bigr) ~=~ \exp(y^2) \erfc(y) ~.
\end{align}
The Voigt function is symmetric, i.e.\ $K(-x,y) = K(x,y)$ and essentially reduces to the Lorentz function for large $|x + \I y|$.
The Voigt function is identical to the real part of the complex function 
\begin{align}\label{wDef}
 w(z) ~&\equiv~ K(x,y) \,+\, \I L(x,y) ~=~ {\I \over \pi} ~ \int_{-\infty}^\infty ~ {e^{-t^2} \over z-t} ~ \D t \\
 ~&=~ \exp(-z^2) \erfc(-\I z)
\qquad\hbox{ with }\quad z = x + \I y.
\end{align}

\subsection{The \citet{Whiting68} approximation}
\label{ssec:Whiting}

To our knowledge the first approximation of the Voigt function using a combination of Lorentz and Gauss functions
\begin{equation}
 \label{pseudo} K(x,y) ~=~ K(0,y) \left[ \bigr( 1-\eta(y) \bigl) G(x) ~+~ \eta(y) L(x) \right] \\
\end{equation}
with 
\begin{align}
 \label{lorentz} L(x)   ~&=~ \left[ 1 + \left({x \over  \xHalf}\right)^2 \right]^{-1}  \\
 \label{gauss}   G(x)   ~&=~ \exp{\left( -\ln 2 (x / \xHalf)^2 \right)}
\end{align}
is due to \citet{Whiting68}.
The weight factor is given by the ratio of the widths of the Lorentz and Voigt profiles or
\begin{align}
\label{whitingWeight} \eta   ~&=~ y / \xHalf \\
\intertext{with the half width}
\label{whitingWidth} \xHalf ~&=~ \half \left(y+\sqrt{y^2 + 4 \ln 2} \right) ~.
\end{align}
An improved approximation is obtained by adding a correction term (i.e.\ $K(x,y) \longrightarrow K(x,y)+C(x,y)$) that is given by ``kind of Lorentzians and Gaussians'',
\begin{align} \label{whitingCorr}
 C(x,y) = 0.016 \, \eta (1-\eta) \, \Biggl[ & \exp{\left(-0.4 (x / \xHalf)^{(9/4)}\right)} \\ \notag
                                           & - {10 \over 10 + (x/\xHalf)^{(9/4)}} \Biggr]
\end{align}
By definition, the approximation \eqref{pseudo} is exact in the center at $x=0$.
Furthermore, both approximations are exact for the limiting cases of pure Lorentz and Gauss functions.
According to the author, ``this approximation matches the Voigt profile within 5 per cent at worst and is generally within 3 per cent or less.''

\subsection{The \citet{Matveev72} approximation}
\label{ssec:Matveev}

The approximation is given by
\begin{equation} \label{matveev}
 K(x,y) ~=~ {\sqrt{\ln 2} \over \xHalf} \: \left[ (1 - \eta) G(x) ~+~ {\eta \over \sqrt{\pi\ln 2}} L(x) \right]
\end{equation}
with a correction term
\begin{align}
\label{matveevCorr}
 C(x,y) ~=&~ {\eta (1 - \eta) \over \sqrt{\pi \ln 2}}  \left({1.5 \over \ln 2} + 1 + \eta \right) \\
          &~~~ \times \Biggl[ 0.066 \, \exp{\left(-0.4 \left({x \over \xHalf}\right)^2 \right)}    \notag \\
          &~~~~~~~~ - {1 \over 40 - 5.5\, \left({x \over \xHalf}\right)^2 + \left(\frac{x}{\xHalf}\right)^4} \Biggr] ~. \notag
\end{align}
The weight $\eta$ is defined as in \eqref{whitingWeight} and the half width is given by a refinement of Whiting's approximation \eqref{whitingWidth}%
\footnote{Note that a prefactor $\gamma_\text{L}$ for the correction term is missing in Matveev's  Eq.\ (5) and is correctly inserted in \citet[][after Eq.\ (12)]{Titov97}.
Furthermore, the factor $\eta$ in \eqref{matveevCorr} is not given in \citet[][Eq.\ (12c)]{Titov97}.}
\begin{align} \label{matveevWidth}
 \xHalf ~&=~ \half \left(y+\sqrt{y^2 + 4 \ln 2} \right) \\ \notag
         &~~~~~~~~ + 0.05 y \left( 1 - {2 y \over y + \sqrt{y^2 + 4 \ln 2}} \right) \\
         &=~ \xHalf^\text{W} + 0.05 y \left( 1 - {y \over \xHalf^\text{W}} \right)
\end{align}
Without correction term \eqref{matveevCorr} \citet{Matveev72} reports a ``greatest error of $~\approx 25\%$ at $\eta=0.1$ and $x\approx 3$''.
With correction the maximum error at the line center does not exceed $0.6\%$, and for $x/\xHalf>6$ the error lies within the limits of $1\%$'' for any $y$. 
In the intermediate frequency regime ``the error nowhere exceeds $3\%$.''

\subsection{The \citet{Kielkopf73} approximation}
\label{ssec:Kielkopf}

Without correction term this approximation is identical to \eqref{pseudo}, but with the weight and half width defined as 
\begin{align}
\label{kielkopfWeight} \eta   ~&=~ {y \xHalf \over 1 + y \xHalf} \\
\label{kielkopfWidth} \xHalf ~&=~ \half y \, \left( 1 + k_\text{e} \ln 2 + \sqrt{(1 -k_\text{e} \ln 2)^2 + {4 \ln 2 \over y^2}} \right)
\end{align}
The correction contains the difference of a Gaussian and Lorentzian multiplied with a rational function of $x$ (see Online \ref{app:constants} for numerical values)
\begin{equation} \label{kielkopfCorr}
 C(x,y) ~=~ \eta (1-\eta) \bigl( G(x) - L(x) \bigr) {k_1 + k_2 x^2 \over 1 + k_3 x^2 + k_4 x^4} ~.
\end{equation}
This approximation ``is accurate to the order of 0.0001 of the peak value of the function''.

\subsection{The \citet{Thompson87} approximation}
\label{ssec:Thompson}

In contrast to the three approximations discussed so far the weight factors of the \citet{Thompson87} (and \citet{Liu01}, next subsection) approaches are defined by power series of $y$.
The pseudo-Voigt function is written as
\begin{equation} \label{pV_Thompson}
 K(x,y) ~=~ {\sqrt{\ln 2} \over \xHalf} \; \left[ (1 - \eta) G(x) + {\eta \over \sqrt{\pi \ln 2}} L(x) \right]
\end{equation}
with the Voigt half width and weight (see Online \ref{app:constants})
\begin{align}
 \label{thompsonWidth} \xHalf ~&=~ \Bigl( t_0 + t_1 y + t_2 y^2 + t_3 y^3 + t_4 y^4 + y^5 \Bigr)^{(1/5)} \\
 \label{thompsonWeight}  \eta ~&=~ \tau_1 {y \over \xHalf}  + \tau_2 \left({y \over \xHalf}\right)^2 + \tau_3 \left({y \over \xHalf}\right)^3
\end{align}
According to \citet{Ida00} the maximum deviation of about $1.2\%$ is found at $y / (y + \sqrt{\ln 2}) \approx 0.5$.

\subsection{The \citet{Liu01} approximation}
\label{ssec:Liu}

Introducing a dimensionless parameter $ d = (y-\sqrt{\ln 2}) / (y+\sqrt{\ln 2})$ and approximating the weights (see \qeq{LiuWeights} in the Online Appendix) as
\begin{align}
 \label{LiuWeightL} c_\text{L} ~&=~ l_0 + l_1 d + l_2 d^2 + l_3 d^3 \\
 \label{LiuWeightG} c_\text{G} ~&=~ g_0 + g_1 d + g_2 d^2 + g_3 d^3
\end{align}
the pseudo-Voigt function is written as
\begin{equation}
 K(x,y) ~=~  {c_\text{L} \over \xHalf \sqrt\pi} L(x) ~+~ c_\text{G} {\sqrt{\ln 2} \over \xHalf} \, G(x)
\end{equation}
with the half width given by the \citet{Olivero77} approximation 
\begin{equation}
\begin{aligned}
 \xHalf(y) ~&=~ \bigl( y + \sqrt{\ln 2} \bigr) \; \bigl( 1 - 0.18121 (1 - d^2) - \beta \sin\pi d \bigr) \\
 \beta     ~&=~ 0.023665 \exp{(0.6 d)} + 0.00418 \exp{(-1.9 d)}
\end{aligned}
\end{equation}
According to the abstract \citep{Liu01} ``the maximum errors of width, area, and peak \dots\ are 0.01\%, 0.2\%, and 0.55\%, respectively.''


\begin{figure}[h]
 \centering\includegraphics[width=\linewidth]{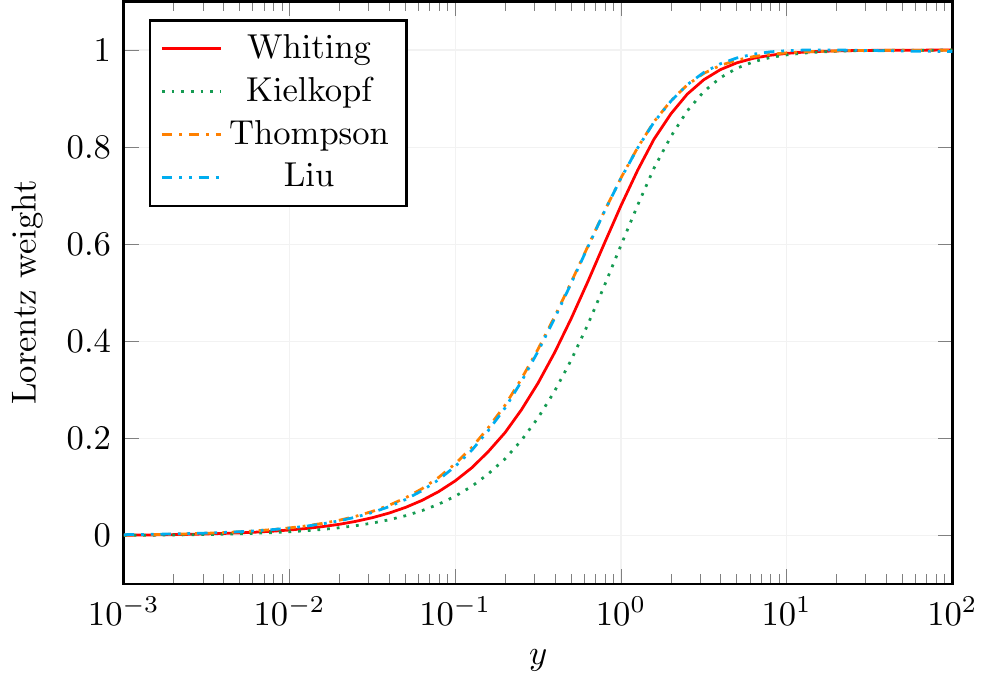}
 \caption{The weight of the Lorentzian: $\eta$ as defined in \eqref{whitingWeight}, \eqref{kielkopfWeight}, \eqref{thompsonWeight} and $c_\text{L}$ defined in \eqref{LiuWeightL}.}
 \label{fig_wgt}
\end{figure}

\section{Results}
\label{sec:results}

Ignoring the correction terms Eqs.\ \eqref{whitingCorr}, \eqref{matveevCorr}, and \eqref{kielkopfCorr}, all algorithms use a weighted sum of the Lorentz and Gauss function,
where the Lorentz weight tends to one for $y\gg 1$ and zero for $y\ll 1$.
\qufig{fig_wgt} compares the weights of the Lorentz function for all five approximations.
The weights of \citeauthor{Whiting68} and \citeauthor{Matveev72} are identical by construction, \qeq{whitingWeight}, and the \citeauthor{Thompson87} and \citeauthor{Liu01} weights appear to be very similar (with differences for small and large $y$ only).
For $y \approx 1$ (i.e.\ equal width of the Lorentz and Gauss function) all weights are approximately 0.7.
Note that except for \citeauthor{Liu01} the Gaussian weight is simply given by $1-\eta$.

\begin{figure*}
 \centering\includegraphics[width=\textwidth]{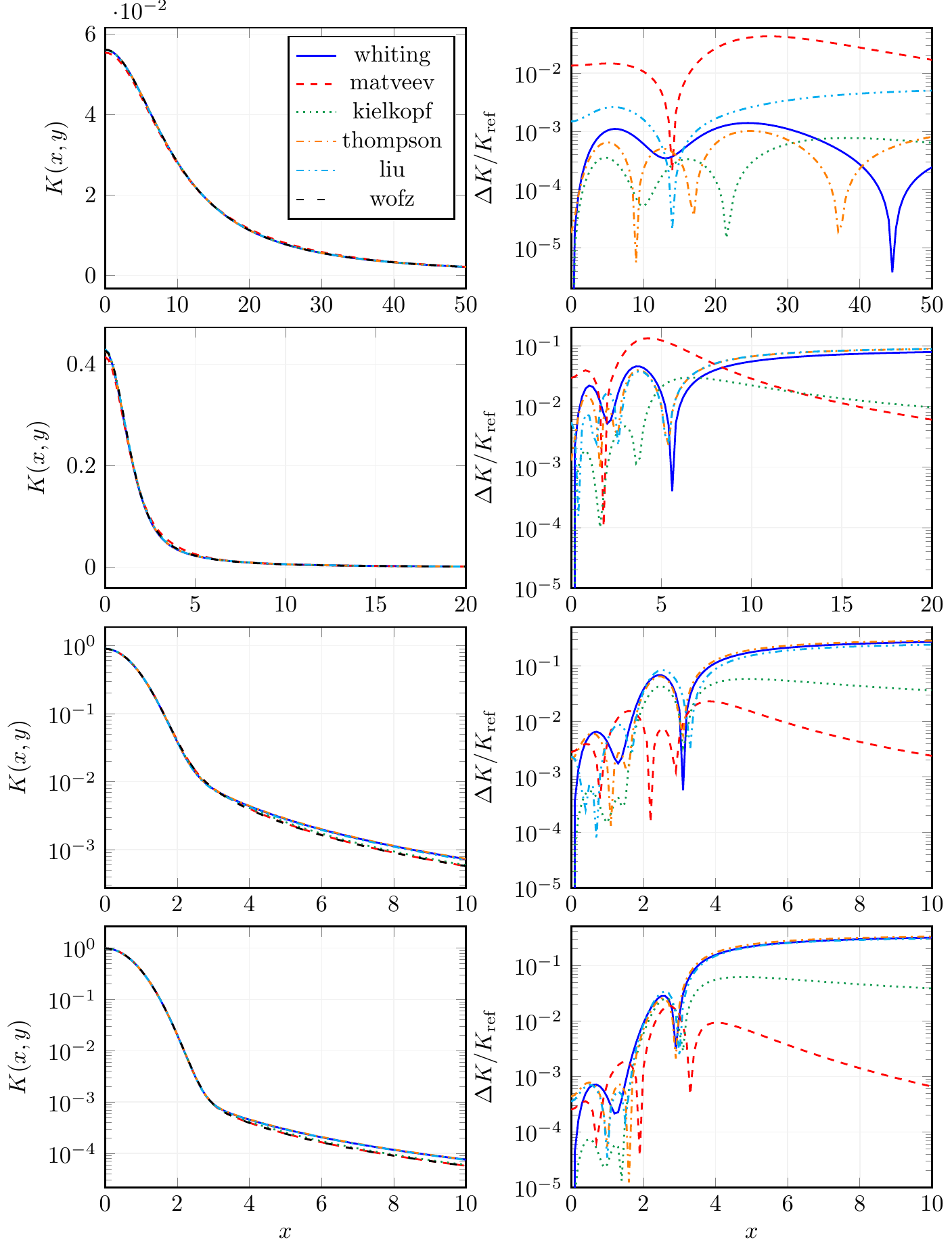}
 \caption{The Voigt function (left) and the relative error (right) for $y=10$ (top), $y=1.0$, $y=0.1$, and $y=0.01$ (bottom).
          Note the different range of $x$ values in the four rows and the linear $y$ axis of the top left plots.}
 \label{figError}
\end{figure*}

As discussed in \citet{Schreier11v} and in our previous assessments of simple closed-form approximations of the Voigt function \citep{Schreier16,Schreier17g,Schreier18v},
the range of $y$ values encountered in molecular spectroscopy and atmospheric and astrophysical applications spans many orders of magnitude.
In \qufig{figError} (left) we compare the pseudo-Voigt approximations with reference values for $w(z)$, Eq.\ \eqref{wDef}, obtained with the \texttt{wofz} code
(algorithm originally based on \citet{Poppe90,Poppe90a} and later refined with ideas from \citet{Zaghloul11};
Scientific Python (\url{http://scipy.org}) implementation \texttt{scipy.special.wofz} with at least 13 significant digits according to the documentation).

The function values shown on the left appear to be in reasonably good agreement with the reference.
However, significant problems show up in the relative errors $|K - K_\text{wofz}| / K_\text{wofz}$ (right side)
and for all approximations the maximum error is larger than 1 percent.

In the line center ($x=0$) the Whiting and Kielkopf approximations are exact by definition (assuming that the exponentially scaled complementary error function $K(0,y) = \erfce(y)$, \qeq{vgtFct0}, is evaluated exactly).
The other approximations have errors of some percent for $y=1$, but smaller errors for smaller and larger $y$.
Note that in the line center the Whiting and Kielkopf corrections vanish, i.e.\ $C(0,y)=0$, in contrast to the Matveev correction.

In the line wings only Matveev's approximation (with and without correction) has relative errors decreasing with increasing $|x|$.
Evaluating the Voigt function for very large $x$ ($x \le 2000$ for $y=10$ and $x \le 100$ for $y=1$) indicates that
for all other approximations the errors become constant for large $x$.
For Kielkopf this asymptotic error is about $10^{-4}$ for $y=10$ and $0.02$ for $y=0.1$ and $0.01$.

For small $y\le 0.1$ Matveev's approximation appears to be superior (with errors less than a few percent), however, for $y>1$ the correction term worsens the accuracy,
and the results shown in the top of \qufig{figError} have been obtained without the correction \eqref{matveevCorr}.
For Kielkopf and Matveev, the largest errors occur for intermediate values of $x$,
whereas the other approximations always fail for large $x$.

\begin{figure*}
 \includegraphics[width=\textwidth]{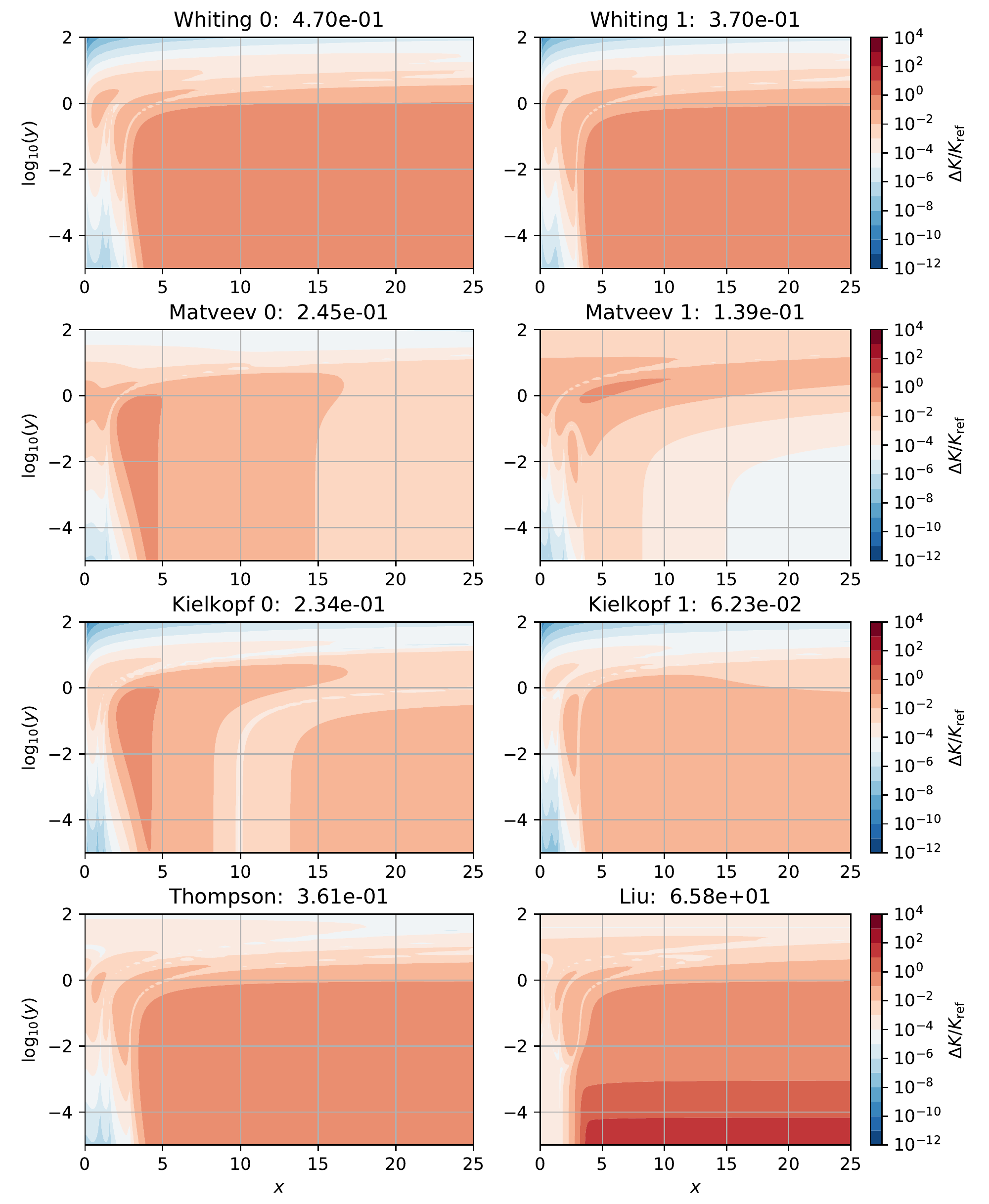}
 \caption{Contour plots of the relative error of the pseudo-Voigt approximations:
          In the first three rows the relative errors of the Whiting, Matveev, and Kielkopf approximations without correction terms are shown on the left, and with correction on the right. The number in the title indicates the maximum relative error.}
 \label{figContour}
\end{figure*}

The contour plots of relative errors shown in \qufig{figContour} essentially confirm these findings.
Except for the \citet{Liu01} Lorentz-Gauss combination all approximations have at least a small region where the relative accuracy is better than $10^{-4}$
(see \citep{Schreier11v} for a discussion of the $10^{-4}$ criterion).
For Whiting and Kielkopf a relative high accuracy is achieved near the origin for small and large $y$, for Matveev and Thompson only for small $y$.

The maximum relative error indicated in the title of all subplots identifies Kielkopf's code (with correction) as the most accurate approximation.
However, this maximum error is relatively large for modern standards (about six percent in the wings),
and for small $y$ Matveev's code appears to be better.

Despite the significant accuracy problems of all pseudo-Voigt approximations it is nevertheless instructive to test the numerical performance of the codes.
Simple tests within the IPython interpreter \citep{Perez07} indicate that the Kielkopf and Matveev approximations are somewhat slower than the optimized combination of
the \citet{Humlicek82} and \citet{Weideman94} rational approximations \citep{Schreier11v}.
However, evaluation of {HNO$_3$} cross sections in the microwave for a series of pressures and temperatures corresponding to Earth's atmosphere in the $ 0 \,\text{--}\, 120\rm\,km$ altitude
range is about a factor two slower with these two pseudo-Voigt approximations compared to the \citeauthor{Humlicek82}--\citeauthor{Weideman94} combination.
For details see the online appendix \ref{app:speed}.


\section{Summary and Conclusions}
\label{sec:conclusions}

Closed-form expressions for the Voigt function based on combinations of Lorentz- and Gauss-type functions show significant accuracy problems,
with relative errors in the percent range.
Note that in this study we have only considered ``pseudo-Voigt'' approximations based on an analytical, closed-form expression for the weight of the Lorentz and Gauss components
(as a function of $y$, the ratio of the Lorentz and Gauss width).
In several studies pseudo-Voigt approximations have been used for analysis of experimental data, where the weight has been estimated by least squares fitting \citep[e.g.][]{Wertheim74,SanchezBajo97}.

Our conclusions now are therefore similar to those given in \citet{Schreier16,Schreier17g}:
Closed-form expressions as presented here might be desirable for certain applications, but their quality is limited.
In general approximations based on modern state-of-the-art numerical methods, e.g.\ rational approximations as discussed in \citet{Humlicek79, Humlicek82, Weideman94} and \citet{Schreier11v,Schreier18h} are recommended.

\medskip 
\section*{Acknowledgments}
Financial support by the Deutsche Forschungsgemeinschaft --- DFG (project SCHR 1125/3-1) is gratefully acknowledged.

\medskip 

\begin{small}
\bibliography{JOURNALS,comp,math,molec,radiation,voigt}
\end{small}

\newpage
\clearpage

\centerline{\fbox{\parbox{\linewidth}{\emph{\LARGE Supplementary Material} \\[1ex]
 Franz Schreier \\[1ex]
 An assessment of some closed-form expressions for the Voigt function III:
 Combinations of the Lorentz and Gauss functions \\[1ex]
 J.\ Quant.\ Spectroscopy \& Radiative Transfer, 2019 \\
 doi: 10.1016/j.jqsrt.2019.01.017
 }}} 

\appendix
\section{Translations}
\noindent
The Lorenz, Gauss, and Voigt profiles are defined as
\begin{align}
 g_\text{L} (\nu-\hat\nu, \gamma_\text{L}) ~=&~ {\gamma_\text{L} / \pi \over (\nu - \hat\nu)^2 + \gamma_\text{L}^2} ~, \\
 g_\text{G} (\nu-\hat\nu, \gamma_\text{G}) ~=&~ {1 \over \gamma_\text{G}} \left({\ln 2 \over \pi}\right)^{1/2} \cdot
                                                 \exp \left[ - \ln 2 \left({\nu - \hat\nu \over \gamma_\text{G}} \right)^2 \right] ~.  \\
 g_\text{V} (\nu-\hat\nu, \gamma_\text{L}, \gamma_\text{G}) ~=&~ {\sqrt{\ln 2 / \pi} \over \gamma_\text{G}} \, K(x,y)
\end{align}
with normalization $\int g(\nu,\dots) \, \mathrm{d}\nu =1$ and half width at half maximum (HWHM) $\gamma$.
The dimensionless variables of the Voigt function $K$ are defined as ratios
\begin{equation}\label{defxy}
 x ~=~ \sqrt{\ln 2}~ {\nu - \hat\nu \over \gamma_\text{G}} 
\qquad\hbox{ and}\quad
 y ~=~ \sqrt{\ln 2}~ {\gamma_\text{L} \over \gamma_\text{G}} ~.
\end{equation}
In the definition of the pseudo-Voigt functions the following ratios of the widths are used frequently
\begin{align}
 {\gamma_\text{L} \over \gamma_\text{V}} ~&=~ {y \over \xHalf} \\
 {\gamma_\text{G} \over \gamma_\text{V}} ~&=~ {\sqrt{\ln 2} \over \xHalf} ~.
\end{align}

\section{Numerical Constants}
\label{app:constants}

\noindent
The constant used in the \citet{Kielkopf73} half width approximation is $k_\text{e} = 0.0990$ and the coefficients of the correction term are
\begin{equation} \label{kielKopfCoeff}
\begin{aligned}
 k_1 ~&=~ +0.8029                      \qquad &k_2 ~&=~ -0.4207 \\
 k_3 ~&=~ +0.2030                     \qquad &k_4 ~&=~ +0.07335 ~.
\end{aligned}
\end{equation}
The coefficients of the half width expansion \eqref{thompsonWidth} of \citet{Thompson87} are
\begin{equation} \label{thompsonWeights}
\begin{aligned}
 t_0 ~&=~ (\ln 2)^{5/2}               \qquad &t_1 ~&=~ (\ln 2)^2 \cdot 2.69269 \\
 t_2 ~&=~ (\ln 2)^{3/2} \cdot 2.42843 \qquad &t_3 ~&=~ (\ln 2) \cdot 4.47163 \\
 t_4 ~&=~ (\ln 2)^{1/2} \cdot 0.07842 \qquad &t_5 ~&=~ 1.0
\end{aligned}
\end{equation}
and the weight expansion \eqref{thompsonWeight} is defined with
\begin{equation}
 \tau_1 = +1.36603, \qquad
 \tau_2 = -0.47719, \qquad
 \tau_3 = +0.11116.
\end{equation}

\noindent
The coefficients of the Lorentz and Gauss weights used by \citet{Liu01} are
\begin{equation} \label{LiuWeights}
\begin{aligned}
 l_0 ~&=~ +0.68188 \qquad &g_0 ~=~ +0.32460 \\
 l_1 ~&=~ +0.61293 \qquad &g_1 ~=~ -0.61825 \\
 l_2 ~&=~ -0.18384 \qquad &g_2 ~=~ +0.17681 \\
 l_3 ~&=~ -0.11568 \qquad &g_3 ~=~ +0.12109 
\end{aligned}
\end{equation}

\begin{figure*}
 \centering\includegraphics[width=0.8\textwidth]{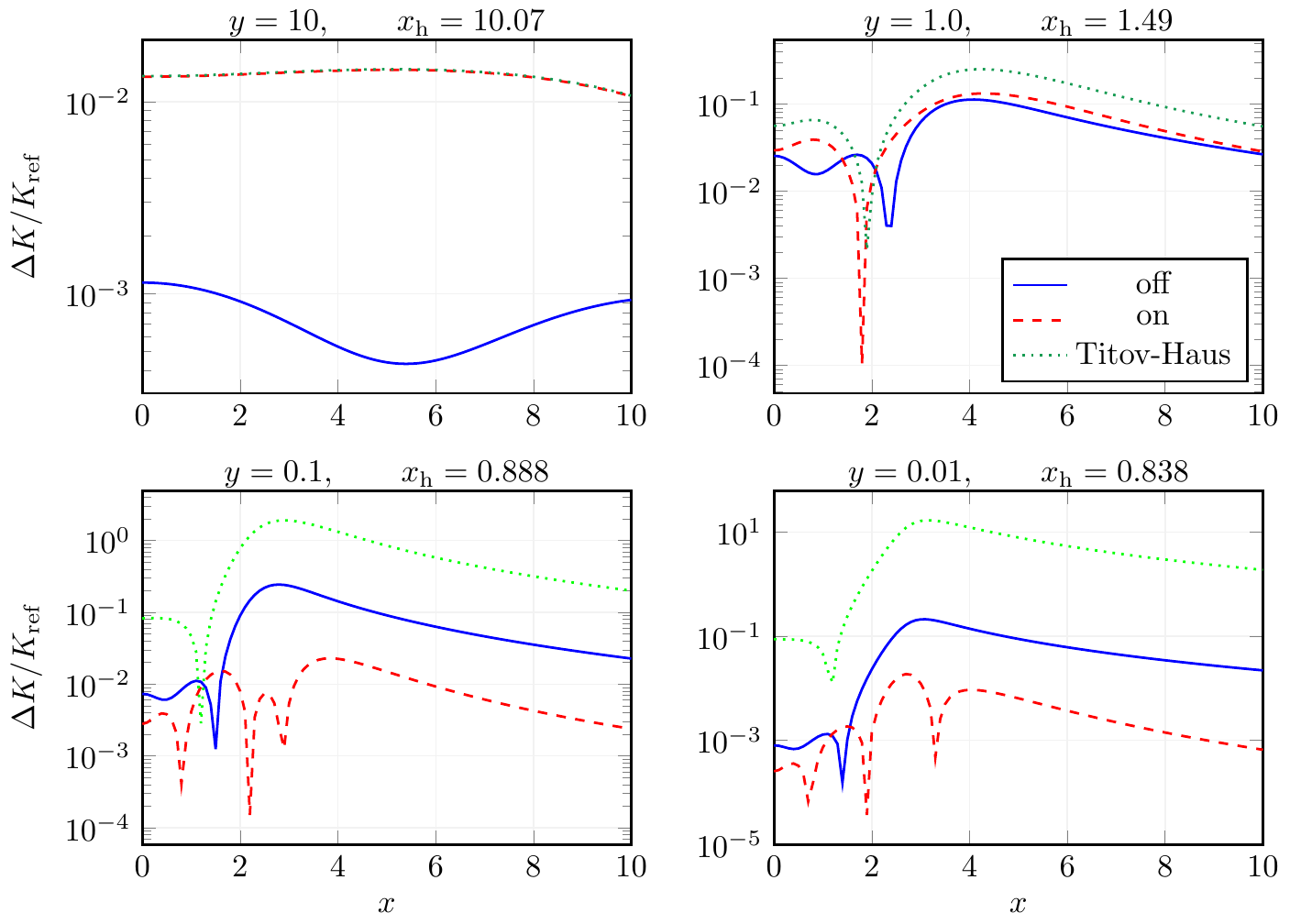}
 \caption{\label{fig:matveev} Comparison of different versions of the \citet{Matveev72} approximation.
         ``off'' and ``on'' indicate Matveev's approximation without/with correction.}
\end{figure*}

\section{Matveev}

As indicated in the footnote of subsection \ref{ssec:Matveev}, some differences show up in the original work by \citet{Matveev72} and in \citet{Titov97}.
Whereas the $y \propto \gamma_\text{L}$ factor in the correction for the width is required for dimension reasons,
our tests depicted in \qufig{fig:matveev} indicate that the weight factor $\eta$ (or $\zeta$ in the original work) has been forgotten by \citet{Titov97}.

\section{Computational efficiency}
\label{app:speed}
For a first, preliminary test of the speed of pseudo-Voigt approximations we have used the IPython \citep{Perez07} builtin ``magic'' function \texttt{\%timeit}:
\\ 
\begin{ttfamily} \small
In [1]: from pseudoVoigt import * \\
In [2]: x=numpy.linspace(0.,100.,10001);  y=1.0 \\
In [3]: \%timeit kielkopf(x,y) \\
424 $\mu$s $\pm$ 273 ns per loop (mean $\pm$ std.\ dev.\ of 7 runs, 1000 loops each) \\
\end{ttfamily}

For the three cases $y=10.0$, $y=1.0$, and $y=0.001$ (see Table \ref{timeit}) the \citeauthor{Humlicek82}--\citeauthor{Weideman94} combination suggested in \citet{Schreier11v} is somewhat faster than the Kielkopf and Matveev approximations.
However, evaluation of one or two exponential(s) (Gaussians) and two fractions for a single $x$ by the Kielkopf and Matveev codes appears to be faster than the numerous multiplications required
for the ``brute-force'' \citet{Weideman94} 24-term rational approximation or the \citet{Humlicek79} rational approximation ``\texttt{zpf16}'' generalized to 16 terms \citep{Schreier18h}.

For a more realistic assessment, molecular cross sections are computed as required for high resolution atmospheric radiative transfer modeling,
i.e. HNO$_3$ cross sections in the $16 \,\text{--}\, 17 \rm\, cm^{-1}$ interval resulting from the superposition of 2376 lines in $6 \,\text{--}\, 27 \rm\, cm^{-1}$ are computed for a series of
pressure and temperature pairs corresponding to altitudes $ 0 \,\text{--}\, 120\rm\,km$ in Earth's atmosphere
(see \citet{Schreier11v,Schreier18h} for more details).
In Python the total time required with the Kielkopf and Matveev approximations is more than a factor two larger than with the \citeauthor{Humlicek82}--\citeauthor{Weideman94} combination.

\begin{table}
 \caption{Execution time (in $\rm \mu s$) measured by the \texttt{\%timeit} function in the IPython interpreter. 
 For all test $0 \le x \le 100$ with $n_x=10001$ grid points.
 The tests have been performed on a desktop with an Intel x86\_64 CPU ``i7-4770'' running at 3.4\,GHz with 8192\,KB cache size.}
 \label{timeit}
 \begin{tabular}{lrrr}
  \hline
  \hfill $y$     &  10.0   & 1.0     & 0.001   \\
  \hline
  Kielkopf       &  459    & 424     & 393     \\
  Matveev        &  451    & 437     & 392     \\
  hum1wei24      &  252    & 326     & 332     \\
  weideman24     &  803    & 801     & 809     \\
  zpf16h         &  693    & 694     & 695     \\
  \hline
 \end{tabular}
\end{table}




\end{document}